\documentclass[twoside,twocolumn,9pt]{article}
\usepackage{extsizes}
\usepackage[super,sort&compress,comma]{natbib} 
\usepackage[version=3]{mhchem}
\usepackage[left=1.5cm, right=1.5cm, top=1.785cm, bottom=2.0cm]{geometry}
\usepackage{balance}
\usepackage{mathptmx}
\usepackage{sectsty}
\usepackage{graphicx} 
\usepackage{lastpage}
\usepackage[format=plain,justification=justified,singlelinecheck=false,font={stretch=1.125,small,sf},labelfont=bf,labelsep=space]{caption}
\usepackage{float}
\usepackage{fancyhdr}
\usepackage{fnpos}
\usepackage[english]{babel}
\addto{\captionsenglish}{%
  
}
\usepackage{array}
\usepackage{droidsans}
\usepackage{charter}
\usepackage[T1]{fontenc}
\usepackage{mathrsfs}
\usepackage[usenames,dvipsnames]{xcolor}
\usepackage{setspace}
\usepackage[compact]{titlesec}
\usepackage{hyperref}

\usepackage{epstopdf}

\definecolor{cream}{RGB}{222,217,201}
\begin{document}

\pagestyle{fancy}
\thispagestyle{plain}
\fancypagestyle{plain}{
\renewcommand{\headrulewidth}{0pt}
}

\makeFNbottom
\makeatletter
\renewcommand\LARGE{\@setfontsize\LARGE{15pt}{17}}
\renewcommand\Large{\@setfontsize\Large{12pt}{14}}
\renewcommand\large{\@setfontsize\large{10pt}{12}}
\renewcommand\footnotesize{\@setfontsize\footnotesize{7pt}{10}}
\renewcommand\scriptsize{\@setfontsize\scriptsize{7pt}{7}}
\makeatother

\renewcommand{\thefootnote}{\fnsymbol{footnote}}
\renewcommand\footnoterule{\vspace*{1pt}%
\color{cream}\hrule width 3.5in height 0.4pt \color{black} \vspace*{5pt}} 
\setcounter{secnumdepth}{5}

\makeatletter 
\renewcommand\@biblabel[1]{#1}            
\renewcommand\@makefntext[1]%
{\noindent\makebox[0pt][r]{\@thefnmark\,}#1}
\makeatother 
\renewcommand{\figurename}{\small{Fig.}~}
\sectionfont{\sffamily\Large}
\subsectionfont{\normalsize}
\subsubsectionfont{\bf}
\setstretch{1.125} 
\setlength{\skip\footins}{0.8cm}
\setlength{\footnotesep}{0.25cm}
\setlength{\jot}{10pt}
\titlespacing*{\section}{0pt}{4pt}{4pt}
\titlespacing*{\subsection}{0pt}{15pt}{1pt}

\fancyfoot{}
\fancyfoot[LO,RE]{\vspace{-7.1pt}\includegraphics[height=9pt]{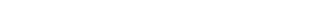}}
\fancyfoot[CO]{\vspace{-7.1pt}\hspace{13.2cm}\includegraphics{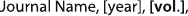}}
\fancyfoot[CE]{\vspace{-7.2pt}\hspace{-14.2cm}\includegraphics{RF}}
\fancyfoot[RO]{\footnotesize{\sffamily{1--\pageref{LastPage} ~\textbar  \hspace{2pt}\thepage}}}
\fancyfoot[LE]{\footnotesize{\sffamily{\thepage~\textbar\hspace{3.45cm} 1--\pageref{LastPage}}}}
\fancyhead{}
\renewcommand{\headrulewidth}{0pt} 
\renewcommand{\footrulewidth}{0pt}
\setlength{\arrayrulewidth}{1pt}
\setlength{\columnsep}{6.5mm}
\setlength\bibsep{1pt}

\makeatletter 
\newlength{\figrulesep} 
\setlength{\figrulesep}{0.5\textfloatsep} 

\newcommand{\topfigrule}{\vspace*{-1pt}%
\noindent{\color{cream}\rule[-\figrulesep]{\columnwidth}{1.5pt}} }

\newcommand{\botfigrule}{\vspace*{-2pt}%
\noindent{\color{cream}\rule[\figrulesep]{\columnwidth}{1.5pt}} }

\newcommand{\dblfigrule}{\vspace*{-1pt}%
\noindent{\color{cream}\rule[-\figrulesep]{\textwidth}{1.5pt}} }

\makeatother

\twocolumn[
  \begin{@twocolumnfalse}
{\includegraphics[height=30pt]{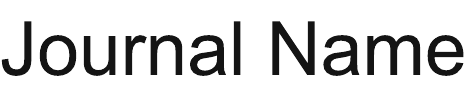}\hfill\raisebox{0pt}[0pt][0pt]{\includegraphics[height=55pt]{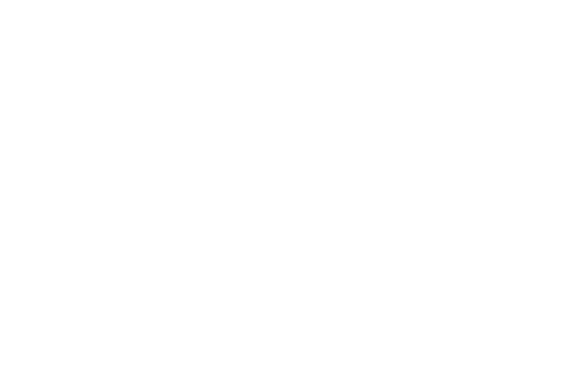}}\\[1ex]
\includegraphics[width=18.5cm]{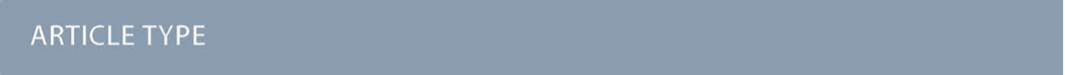}}\par
\vspace{1em}
\sffamily
\begin{tabular}{m{4.5cm} p{13.5cm} }

\includegraphics{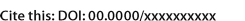} & 
\noindent\LARGE{\textbf{Rectification of stress by fiber networks: Manifestation of non-linear screening through self-organized buckling }} \\
 & \vspace{0.3cm} \\

 & \noindent\large{Kanaya Malakar$^{a,b}$, Albert Countryman$^{a,c}$, and Bulbul Chakraborty$^{a,\dag}$} \\

\includegraphics{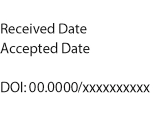} & \\

\end{tabular}

 \end{@twocolumnfalse} \vspace{0.6cm}

  ]

\renewcommand*\rmdefault{bch}\normalfont\upshape
\rmfamily
\section*{}
\vspace{-1cm}


\footnotetext{\textit{$^{a}$~Martin A. Fisher School of Physics, Brandeis University, Waltham, USA.}}
\footnotetext{\textit{$^{b}$~Center for Computational and Integrative Biology, Rutgers University, Camden, USA.}}
\footnotetext{\textit{$^{c}$~Department of Physics and Astronomy, UCLA, Los Angeles, USA. }}
\footnotetext{\textit{$^{\dag}$~E-mail: bulbul@brandeis.edu}}





\sffamily{\textbf{Force transmission at large length scales is crucial for such biological functions as cell motility and morphogenesis. The networks that transmit these forces are malleable, patterned by active forces generated at the microscale by biological motors.  In this paper we explore a simple model of a non-linear fiber network which has only two modes of deformation, but exhibits diverse mechanical phases with distinct large-scale response, tuned by the strength of a microscopic force dipole. We demonstrate, via numerical simulations, that the network is remodeled by organized patterns of buckling, which lead to a renormalization of the Poisson ratio. Finally, we show that the emergent behavior at large length scales can be ascribed to ``mechanical screening'' of the force dipole, analogous to dielectric screening of charges in electrostatics.
}}\\


\rmfamily 



\begin{figure}[h!]
 \centering
 \includegraphics[width = 0.45\textwidth]{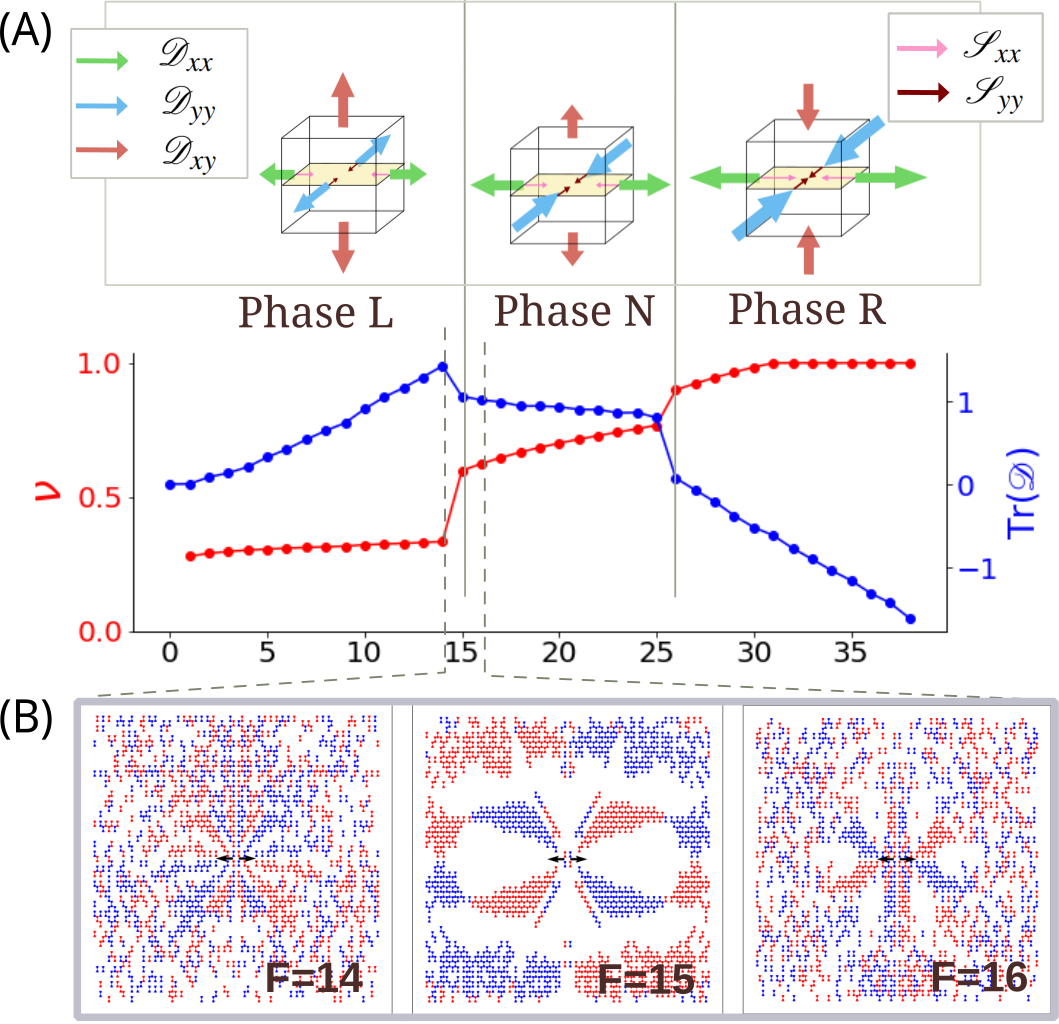}
 \caption{(A) (Top Row) $\mathscr{S}$, and $\mathscr{D}$, visualized using the Voigt notation (see text). A cube represents the three components:$\lbrace \mathscr{D}_{xx},\mathscr{D}_{yy},\mathscr{D}_{xy}\rbrace$), and the mid-plane represents:$\lbrace \mathscr{S}_{xx},\mathscr{S}_{yy},\mathscr{S}_{xy}=0\rbrace$).  Arrows pointing outward (inward) represent positive (negative) stress. (Second Row)  Annular pressure (Blue curve:) ($Tr(\mathscr{D}$)), and Poisson ratio (Red curve), $\nu$,  
exhibit three distinct mechanical phases:  - linear (L), non-linear (N)  and rectified (R) .
 (B) Buckling pattern is shown at three consecutive values of applied force ($F=14, 15, 16$). Ordered domain patterns signal the discontinous change at $|F|=15$-there are no domains in the network  preceding or following these phase changes. Similar trend is seen at $|F|=25$ (not shown).}
 \label{first_image}
\end{figure}
Many fundamental cellular processes require exquisitely orchestrated large-scale reorganization of structural filaments. One mechanism of reorganization is via internal, active forces, generated by motor proteins. The transmission of these forces, mediated by a  non-linear network of fiber-like filaments, is highly tunable~\cite{Madan_hydro1,Madan_hydro2,gagnon_shear-induced_2020,broedersz_molecular_2011,broedersz_modeling_2014,furthauer_design_2021,wyse_jackson_structural_2022, Xu_PRE_2015, Murrell_PNAS_2012, Wang_BioPhysJ_2014, Najma_PNAS_2024, Zakharov_Sc_2024, farhadi_actin_2020, burla_stress_2019, furthauer_self-straining_2019}. 
A biomimetic system composed of isotropic, active, microtubule-bundle networks, for example, exhibits shear-induced gelation~\cite{gagnon_shear-induced_2020}. The gel self-yields due to  internal activity - competition between internal activity-rate and shear-rate leading to a nonmonotonic behavior of  the viscosity~\cite{gagnon_shear-induced_2020, tanaka_viscoelastic_1992}. 

Studies of minimalist models of  nonlinear, active, networks are appealing because they can illuminate the essential physics underlying the complexity of the mechanical response. 
The model that we study here was proposed by Ronceray et al~\cite{ronceray_fiber_2016} , and exhibits \textit{rectification} of the far-field stress under the activity of a local, extensile  force dipole.  In this work, we show that the rectfication is a manifestation of  ``nonlinear elastic screening'' of the force dipole~\cite{nampoothiri_tensor_2022},which emerges from  self-organized buckling patterns.  We trace the microscopic origin of this organized response to soft modes of  an underlying Kagome lattice~\cite{lubensky_phonons_2015,mao_maxwell_2018}.  

\begin{figure}[h]
 \centering
 \includegraphics[width = 0.3\textwidth]{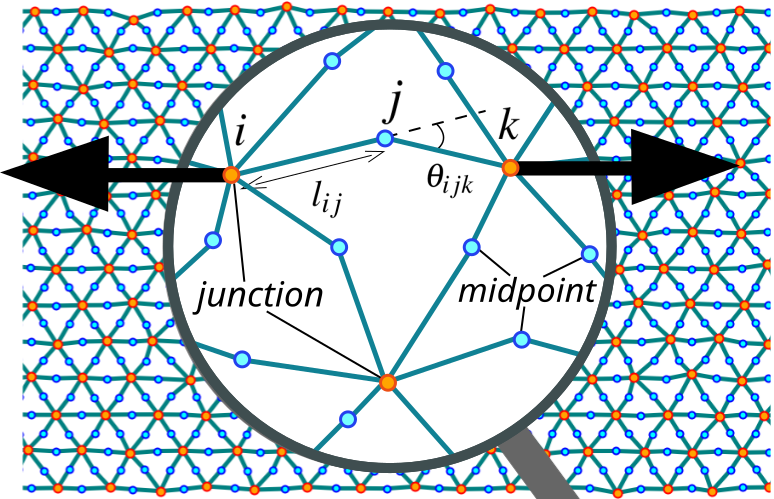}
 \caption{A schematic of the simulated network of non-linear springs. Each spring has two junctions (orange circles) where it is connected to other springs and a midpoint (blue circles) where it can buckle. Each spring can deform either by stretching and contracting or by buckling at its midpoint. Large black arrows indicate the applied force dipole. {The network contains 2900 junctions (nodes)  and 8700 springs.}}
 \label{fig:model}
\end{figure}

We represent an isotropic elastic medium by a regular triangular lattice of non-linear springs. 
Each spring is allowed to buckle at the midpoint as shown in Fig-\ref{fig:model}. Unlike in Ref.~\cite{ronceray_fiber_2016}, we impose periodic boundary conditions, and we do not energetically penalize bending at the junctions of multiple springs.
Two equal and opposite forces, $\bf{F}$  constituting an extensile force dipole, act on two neighboring junctions of springs such that the net force and torque on the entire system is zero (Fig. \ref{fig:model}). 

The total energy of the network is:
\begin{align}
        H = \sum_{\text{segment }(i,j)} \frac{\mu}{2}(l_{ij}-l_0)^2 + \sum_{\text{hinge }(i,j,k)} \kappa \sin^2 \frac{\theta_{ijk}}{2} - \sum_{\text{force }i} {\bf F}_i {\bf r}_i ~,
        \label{eq:model}
\end{align}
where $l_{ij}$ denotes the length of the spring segment $\bf{l}_{ij}$,  $l_0$ is half of the equilibrium length of the spring ${ijk}$, and $\theta_{ijk}$ is the buckling angle at the midpoint $j$. The energy scale of stretching is given by
$\mu$, and $\kappa$ defines the buckling energy scale. 
We fix $\kappa/\mu  = 1/1000$, such that 
buckling is the preferred mode of deformation of the springs.
Details of the simulation are presented in~\cite{Supplementary}.


A useful spatial measure of the nonlinear  network response is provided by the force-moment tensor, $\hat{\sigma}_i = \sum_{ij} {\bf{f}}_{ij} \otimes {\bf{l}}_{ij}$ associated with each node, $i$ of the network, where $\bf{f}_{ij}$ is the force exerted on a node by a segment . The sum is over all spring segments $ij$  connected to the node $i$. {The force moment tensor, coarse-grained over a region,  is an extensive quantity, related to the stress tensor by the volume of the coarse-grained region. For ease of discussion, we refer to $\hat{\sigma}$ as the stress tensor. 

As in Refs. ~\cite{ronceray_fiber_2016,Benoit_Lenz_2023}, we characterize the ``far-field'' response by the tensor,  $\hat{\mathcal{D}} = \sum_{i\in {\rm {boundary}}}({\hat{\sigma}}_{i})$, where the sum is taken over all sites at a prescribed boundary.   In our system with periodic boundary conditions, we measure $\mathscr{D} $ by averaging over all nodes $i$ in an annulus of thickness $\delta{R}= 1$ at a distance of $R=20$ from the force dipole. All lengths are measured in units of the equilibrium length, $l_0$. In these units the dipole strength is $2|F|$.      
    $\hat{\mathcal{D}}$ measures the surface force-dipole moment~\cite{Ronceray_Lenz_2015}.  We also define the total stress in the system as $\hat{\mathcal{S}} = \sum_i \hat{\sigma_i}$. As seen from Fig. \ref{first_image} A, the different ``phases'' or regimes of mechanical response can be characterized by $\lbrace \hat{\mathscr{D}},\hat{\mathscr{S}} \rbrace$. We will show  below that a rigorous mapping of prestress elasticity to a generalized ``electrostatics'' of vector charges (VCT), relates these two quantities~\cite{Ronceray_Lenz_2015,Benoit_Lenz_2023}.  }


Fig. \ref{first_image} provides a summary of the nonlinear response of the network. As demonstrated by Fig. \ref{first_image} A, there are three distinct mechanical phases: 

\begin{enumerate}
    \item phase L: a linear response regime between $|F|=0$ and $|F|\sim 15$, with $Tr(\hat{\mathscr{D}} )$ increasing linearly with $|F|$, and all components of $\hat{\mathscr{D}} $ having the same sign as that of $\hat{\mathcal{S}}$;
    \item phase N: a nonlinear regime with its onset marked by a sharp drop, in $Tr(\hat{\mathscr{D}} )$ at $F=15$, followed by an approximately constant value. This regime is characterized, notably, by the sign of $\mathcal{D}_{yy}$  being opposite to that of $\mathcal{S}_{yy}$ (The force dipole is in the $x$ direction);
    \item phase R: where $Tr(\hat{\mathscr{D}} )$ is negative, indicating a contractile response to an extensile dipole. In this regime, {\it both} $\mathcal{D}_{yy}$ and $\mathcal{D}_{xy}$ have signs opposite to their counterparts in $\hat{\mathcal{S}}$. This phenomenon has been termed {\it rectification}.
\end{enumerate}



\begin{figure}[h!]
 \centering
 \includegraphics[width = 0.45\textwidth]{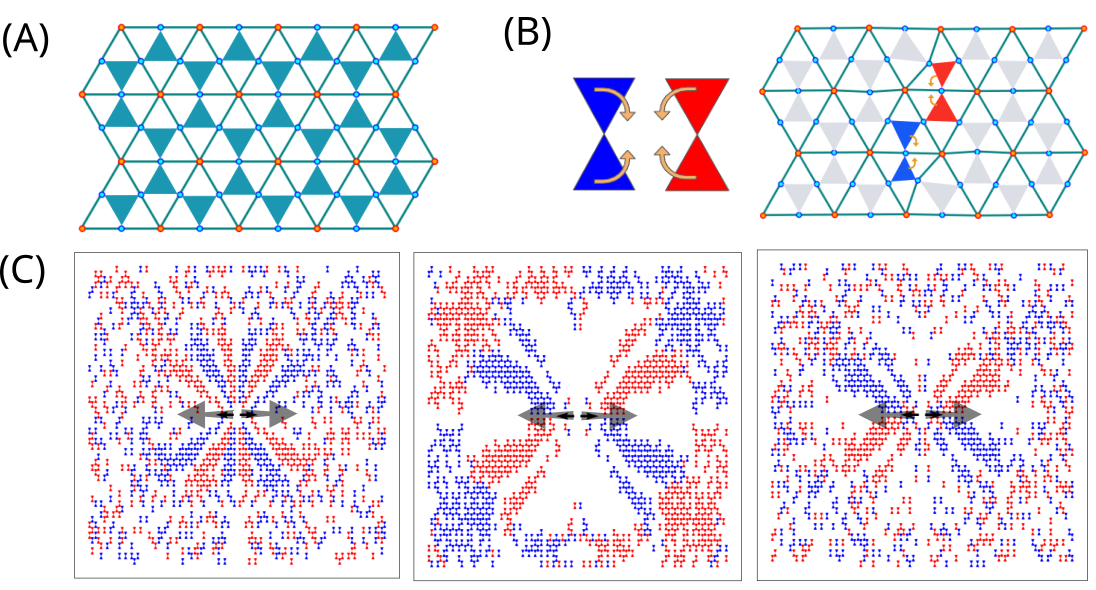}
 \caption{(A) Schematic of a triangular network drawn in green.
The circles denote junctions of the springs  (orange) and  midpoint of each spring (cyan). 
Connecting each midpoint with its four nearest-midpoint-neighbors gives rise to the Kagome lattice shown in teal. (B) under deformation,  a pair of adjacent up and down triangles may rotate in two opposite directions forming a \emph{twisted unit}:colored blue or red according to their rotation direction. Non-twisted units are not colored. (C) Spatial distribution of twisted units at three consecutive force values $F=25$ (left), $F=26$ (center), and $F=27$ (right). The center panel depicts the organized pattern at  the transition from N to R.}
 \label{kagome}
\end{figure}
As shown in Fig. \ref{first_image} B, the spatial pattern of buckling evolves across the boundaries between mechanical phases. 
Figure \ref{kagome}A shows that the network of midpoints forms a Kagome lattice.
The Kagome lattice is well-known for having soft modes~\cite{mao_soft_2010,mao_maxwell_2018}.  
In~\cite{Supplementary}, we present details of a construction based on this Kagome network, that reveals clear signatures of organized buckling. A pair of up and down triangles with opposite chirality form a twisted unit as shown in Fig \ref{kagome}B, which are colored according to the chiralities of the component triangles.
The spatial distribution of these twisted units show two distinct patterns (i) a diffused state, where the twisted units are distributed more or less uniformly throughout the system, and (ii) an ordered state where the twisted units self-organize to form discrete large domains.
We find that  the emergence of these ordered domain structures  is strongly correlated with the stress response characterizing the mechanical phases.
This is illustrated in Fig.  \ref{kagome}C, which show the spatial distribution of twisted units at three consecutive force values ($|F|=25, 26, 27$) chosen such that the system on the left ($|F|=25$) belongs to phase N, the one in the middle ($|F|=26$) sits right at the boundary between phase (N) and phase (R), and the system on the right ($F=27$) belongs to phase R, as shown in Fig. \ref{first_image} A.
These images show  that when the system is well inside a  phase, either (N) or (R), the domains of twisted regions (blue or red) are much smaller than the domains  
observed at the boundary between (N) and (R), as shown in the central panel at $|F|=26$.
This is observed also at the boundary between phases (L) and (N), and is shown in Fig. \ref{first_image}B. 
The left panel in Fig. \ref{first_image}B is in phase (L), the middle panel is at the boundary and the rightmost panel is in phase (N). 

The above analysis demonstrates a correlation between changes in the buckling patterns and transitions between the mechanical phases, defined by the far-field stress response.
The spatial field, $\hat{\sigma}_i$, represents   the prestress, or residual stress, that is stored in the network ``in response to the forcing''.  This stress is not annealed away during the energy minimization process, and arises in {\it response} to the extensile force dipole:  in its absence ($|F|=0$), all springs can relax to their equilibrium length, and $\hat{\sigma}_i=0$ for all nodes.   The nonlinear response, and in particular, the rectified, far-field  response,  has its origin in this prestress.
\begin{figure}[h!]
 \centering
 \includegraphics[width = 0.45\textwidth]{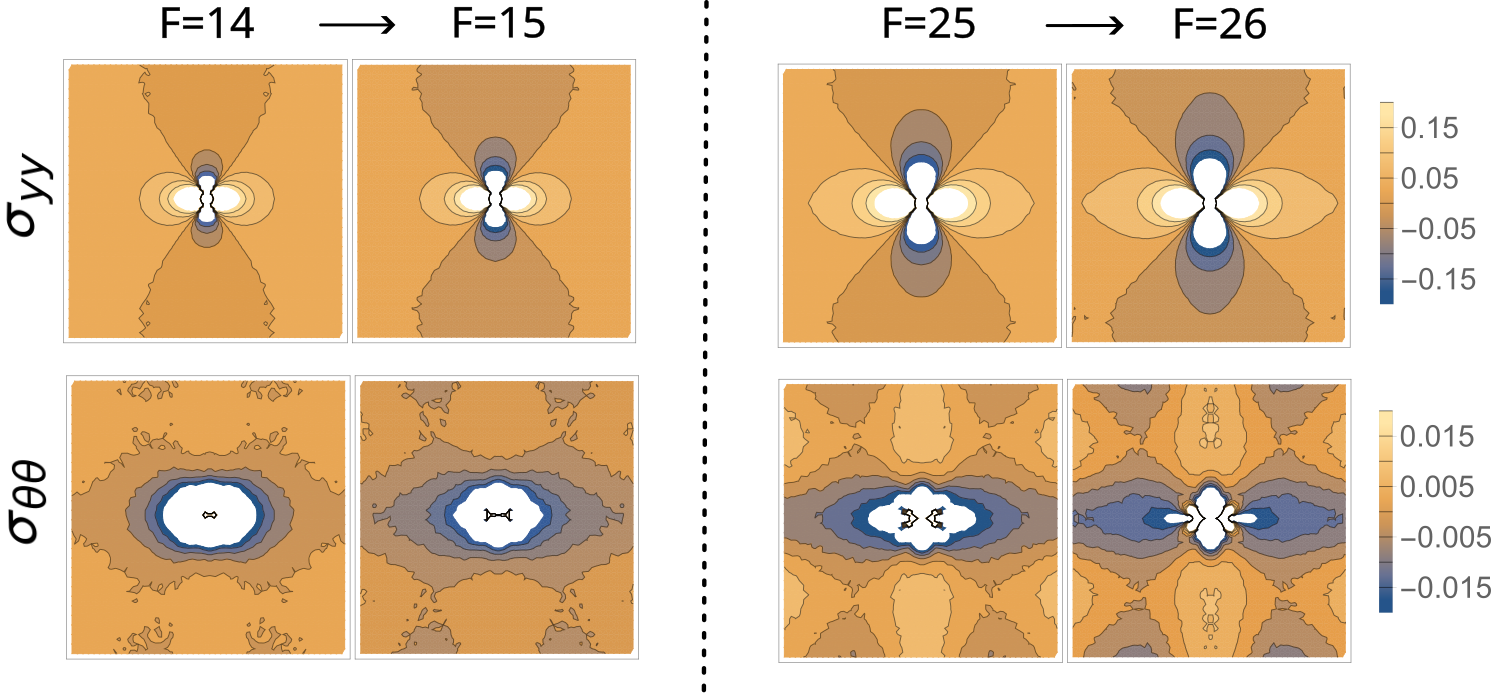}
 \caption{{ Top Panel: Spatial pattern of $\sigma_{yy}$ as we move from phase $L \rightarrow$ phase $N$ (left) and phase $N \rightarrow$ phase $R$ (right). Bottom Panel: Corresponding patterns of $\sigma_{\theta \theta}.$ Fig \ref{first_image}B and fig \ref{kagome}C show the changes in the kagome patterns at these force values.}}
 \label{fig:transitions}
 \end{figure}
Fig. \ref{fig:transitions}, shows the simulation results for the prestress field, $\hat{\sigma}_i$. 
It is clear from this figure that $\sigma_{\theta,\theta}(x,y)$, and $\sigma_{yy} (x,y)$, show distinctive anisotropic spatial patterns in the three mechanical phases, (L), (N) and (R). Comparison with the buckling patterns,  also shown in Figs. \ref{fig:transitions}, and in Figs. \ref{first_image}, and \ref{kagome},  indicates that the prestress pattern reflects the changes in anisotropy of the buckling patterns.  This is most clearly demonstrated by the changes in $\sigma_{\theta \theta}$.

Rectification is observed only in the non-linear elastic network which has buckling modes.
This suggests a mechanism of ``screening'' via buckling deformations such that  the far-away points are not able to ``{see}'' the true nature of the applied force dipole.
This phenomenon is reminiscent of the screening of external charges by a polarizable medium (a dielectric)  in electrostatics~\cite{Zangwill_EandM}.
VCT is an analogous theory of  ``mechanical screening'' of external forces,  which has been shown to describe the mechanical response of solids with prestress~\cite{surajit_arxiv,nampoothiri_tensor_2022}.  
A crucial feature of this formalism is that the stress-bearing network determines the polarizability, and the network in turn is shaped by the polarization tensor.  
Below, we demonstrate that a nonlinear, mechanical screening mechanism is induced through the buckling springs and is the origin of the  observed rectification. 
The complete set of equations representing the continuum theory of  prestressed elasticity~\cite{nampoothiri_tensor_2022} for our system are:
\begin{align}
     \partial_{\alpha} \sigma_{\alpha \beta} = F_{\beta} &~ ~~(\rm {VCT ~Gauss's ~Law}),\nonumber\\
     \partial_{\alpha} E_{\alpha \beta} = F_{\beta}+f_{\beta} &, \quad \partial_{\alpha} P_{\alpha \beta} = -f_{\beta}\nonumber\\
     \hat{\sigma} = \hat{E} + \hat{P},\quad P_{\alpha \beta} &= \chi_{\alpha \beta \gamma \delta} E_{\gamma \delta},\nonumber\\
     E_{\alpha \beta}=\frac{1}{2}(\partial_{\alpha}\varphi_\beta+\partial_{\beta}\varphi_{\alpha}) &\implies \epsilon_{\alpha \beta} \epsilon_{\gamma \delta} \partial_{\alpha} \partial_{\beta} E_{\gamma \delta}=0,\nonumber\\
     \sigma_{{\alpha \beta}} = (\delta_{\alpha \beta \gamma \delta} + \chi_{\alpha \beta \gamma \delta})& E_{\gamma \delta} \equiv {K_{\alpha \beta \gamma \delta}E_{\gamma \delta}} ~,
    \label{eq:VCTG_field_eqns}
\end{align}
where $\bf{f}$, arise from the extension, compression, and buckling of the springs, and $\bf{F}$ is the dipole force appearing in Eq. \ref{eq:model}. 
 The physical observables in this  mechanical screening framework can be mapped to that of an electrostatic dielectric~\cite{Zangwill_EandM} as follows:\\
 \phantom{mediumgap}$\hat{E} \Leftrightarrow$ unscreened electric field\\
 \phantom{mediumgap}$\hat{\sigma} \Leftrightarrow$ screened electric displacement field\\
 \phantom{mediumgap}$\hat{P} \Leftrightarrow$ polarization field.\\
As written originally, VCT-elasticity is a linear dielectric theory, with $\hat P \propto \hat{E}$ via a linear dielectric susceptibility tensor, $\chi$.  The fourth rank tensor $K$ is the analog of  the dielectric tensor in electrostatics, and  plays the role of the elastic modulus tensor in this stress-only formulation of elasticity~\cite{nampoothiri_tensor_2022,nampoothiri_emergent_2020}.

The macroscopic measures used to characterize the different mechanical phases, can be understood in terms of these fields.  $\hat{\mathcal{D}} $ represents the surface force dipole, computed from $\bf{f}$ on all springs linked to sites on the annulus at $R$.  The spring forces, are sources of the induced polarization, $\hat{P}$. Recall that in the absence of the extensile force dipole, $\bf{f} \equiv 0$ on all segments. Therefore~\cite{Ronceray_Lenz_2015}, $${\mathcal{D}}_{\alpha \beta}\equiv \sum_{ij \in {\rm boundary}} (f_{ij})_{\alpha}   (l_{ij})_{\beta} = -\sum_{ij \in {\rm boundary}} \hat{n}_{\lambda} P_{\alpha \lambda} r_{\beta} ~,$$ is a  measure of the induced polarization, as for example, in a capacitor with a dielectric between the plates. $\hat{\mathcal{S}}$, on the other hand, is the integral of $\hat{\sigma}$, and measures the total stress in the system. In ~\cite{Ronceray_Lenz_2015}, these two quantities have been related to the force-moment of the imposed dipole. The VCT relation between $\hat{\sigma}$, and $\hat{P}$, in Eq. \ref{eq:VCTG_field_eqns} shows that they are related via the tensor $K$:  $\hat{\sigma} = (\hat{I} + (\hat{\chi})^{-1}) \hat{P}$, within the linear theory,  the emergent elastic moduli, $K$, relates the two macroscopic measures, $\lbrace \hat{\mathcal{D}}, \hat{\mathcal{S}} \rbrace$.
As we demonstrate in the following (see also Fig. 1), the Poisson ratio is renormalized, strongly, by the embedded force-dipole.

%
This linear dielectric framework can be readily generalized to a nonlinear theory by making $\chi$ a function of $\hat{E}$:
\begin{align}
    P_{\alpha \beta} = \chi_{\alpha \beta \gamma \delta} (\hat{E}) E_{\gamma \delta}~, \quad 
    \sigma_{\alpha \beta} = K_{\alpha \beta \gamma \delta} (\hat{E}) E_{\gamma \delta} ~, 
\end{align}
where the field $\hat E$ can be modified by external forces, thus leading to effective elastic moduli, $K$, which depend on the external forces (non-linear response). In the context of this work, $K$ is a function of the magnitude of the force dipole, $|F|$ . In this nonlinear theory, the relation between $\hat{\sigma}$ and $\hat{P}$, is not linear, and cannot be solved analytically, in general.
Assuming an isotropic form for $K$, mechanical reponse is characterized by two parameters, $\mu $ (mapping to the the shear modulus) and $\nu$ (mapping to the Poisson ratio). In general,  VCT can be used to predict the stress response in terms of $K$.  In 2D, the response depends only on $\nu$~\cite{nampoothiri_tensor_2022}. 
 \begin{figure}[h!]
 \centering
\includegraphics[width = 0.35\textwidth]{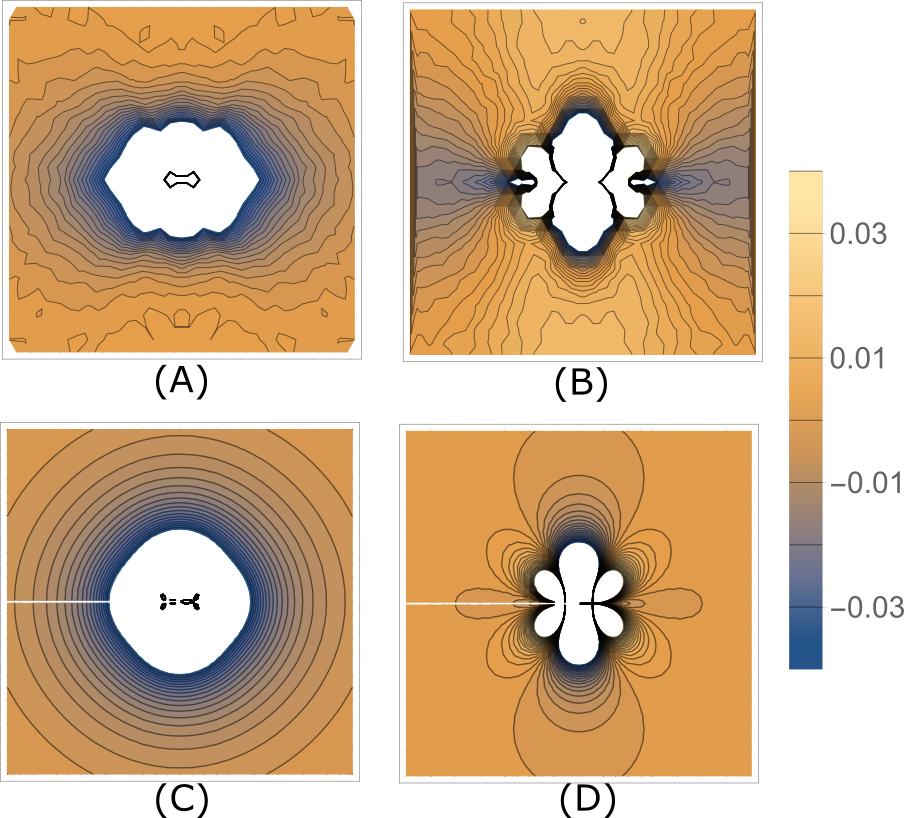}
 \caption{$\sigma_{\theta \theta}$ calculated from simulation at (A) F=10 and (B) F=30. Predictions of $\sigma_{\theta \theta}$ from VCTG where the value of $\nu$ is obtained by fitting the data to theory: (C) $\nu$ = 0.34 and (D) $\nu$=1.0~.}
 \label{fig:stressTT}
\end{figure}
{ The VCT predictions~\cite{Supplementary} for the stress response, in terms of $\nu(|F|)$ are: \\
$\tilde{\sigma}_{xx} = \mathcal{A} q_x(q_x^2+q_y^2(2+\nu))$, 
$\tilde{\sigma}_{yy} = \mathcal{A} q_x(-q_y^2+q_x^2\nu) $, and 
$\tilde{\sigma}_{xy} = \mathcal{A} q_y(q_y^2+q_x^2\nu)$, 
where $\mathcal{A} = -2|F|\sin(dq_x/2)/q^4$.
We have suppressed the argument of $\nu$ to simplify notation.  Inverse Fourier transforms of these expressions yield the real-space fields, $\sigma_{\alpha \beta} (x,y)$. We note that in Fig. \ref{first_image}, we have used the Voigt notation to represent a symmetric second rank tensor, such as $\hat{\sigma}$ as a vector: $\lbrace \sigma_{xx},\sigma_{yy},\sigma_{xy} \rbrace$.}

We extract  
the Poisson ratio ($\nu (|F|)$) by fitting the simulation results for the stress response to the theoretical predictions.
Table \ref{nonlinear} shows the values of the Poisson ratio as a function of $|F|$. 
\begin{table}[h]
 \begin{center}
  \begin{tabular}{ |c|c|c|c|c|c|c| }
     \hline
     F& 5 & 10 & 15 & 20 & 25 & 30 \\\hline 
     $\nu$  & 0.33 & 0.34 & 0.62 & 0.72 & 0.90 & $\approx 1$   \\
     \hline
  \end{tabular}\caption{\footnotesize{Table of Poisson ratios obtained by fitting stress response to the VCTG theory.}}
  \label{nonlinear}
 \end{center}
\end{table}
We find that for $F<15$ the Poisson ratio is close to $1/3$ which is the value of $\nu$ for a linear triangular network. 
Beyond $F=15$, it increases all the way up to a value of $\approx 1$ which is the upper limit for Poisson ratio in two-dimensions \cite{ting_poissons_2005} \cite{shivers_nonlinear_2020}. Our best fit estimate at $|F|=30$ is $1.0$. Figure \ref{first_image}A show the dependence of Poisson ratio on the applied force.
Doing the same fitting exercise with data from the linear network yields a Poisson ratio of $1/3$ across the board.  { In fig \ref{fig:stressTT} , we test the validity of this approach by comparing the stress responses predicted using the fitted values of $\nu (|F|)$ to the simulated ones~\cite{Supplementary}.}
 \\



The analysis we have presented, demonstrates that a nonlinear spring network can reorganize to screen the effects of imposed forces, leading to far-field responses that can entirely mask the imposed force. In particular, this response would indicate no forcing (complete screening) in phase N, and it would indicate the sign of the perturbing force as being opposite to the actual one in phase R (overscreening). These responses, which are ``anomalous'' within a classical elasticity framework, are natural outcomes within the mechanical screening perspective presented in this work. { These three phases, correspond to the phases discussed in Fig. 3 of ~\cite{Benoit_Lenz_2023}, where the analogs of $\hat{\mathcal{D}} $, and $\hat{\mathcal{S}}$, are computed explicitly, in the limit of small prestress.  In our work, we have used the VCT predictions of stress-response to deduce the renormalization   of the  Poisson ratio by the imposed force dipole. We have traced the microscopic origin to the organized buckling patterns for this particular model. The conclusion that reorganization of the stress-bearing network (here through buckling) can screen force dipoles is much more general. It would be illuminating to apply this screening picture to dynamically adapting networks in response to active forcing in biological networks such as the cytoskeleton.One major advantage of the VCT approach is that the notions of strain and a reference state, which are difficult to define in adaptable, renewable force-bearing networks, are not required for analyzing mechanical response.}

\section*{Acknowledgements}
We acknowledge fruitful discussions with Danny Hellstein, John Berezney, and Kabir Ramola. This work was supported by grants NSF MRSEC DMR 2011846, NSF-DMR-2026834, and NSF CBET-2228681. Part of this work was carried out at the Kavli Institute for Theoretical Physics (KITP), supported by grant NSF PHY-2309135.

\section*{Author Contributions}
BC and KM conceptualized the project. KM designed the simulation. KM and AC acquired data. All three authors analyzed results and wrote the manuscript. BC acquired funding for the project.

\section*{Conflicts of interest}
There are no conflicts to declare.

\section*{Data availability}

Data are available upon request from the authors.







\bibliography{MyLibrary,bulbul_buckling} 

\providecommand*{\mcitethebibliography}{\thebibliography}
\csname @ifundefined\endcsname{endmcitethebibliography}
{\let\endmcitethebibliography\endthebibliography}{}
\begin{mcitethebibliography}{30}
\providecommand*{\natexlab}[1]{#1}
\providecommand*{\mciteSetBstSublistMode}[1]{}
\providecommand*{\mciteSetBstMaxWidthForm}[2]{}
\providecommand*{\mciteBstWouldAddEndPuncttrue}
  {\def\EndOfBibitem{\unskip.}}
\providecommand*{\mciteBstWouldAddEndPunctfalse}
  {\let\EndOfBibitem\relax}
\providecommand*{\mciteSetBstMidEndSepPunct}[3]{}
\providecommand*{\mciteSetBstSublistLabelBeginEnd}[3]{}
\providecommand*{\EndOfBibitem}{}
\mciteSetBstSublistMode{f}
\mciteSetBstMaxWidthForm{subitem}
{(\emph{\alph{mcitesubitemcount}})}
\mciteSetBstSublistLabelBeginEnd{\mcitemaxwidthsubitemform\space}
{\relax}{\relax}

\bibitem[Abhishek \emph{et~al.}(2025)Abhishek, Dhanuka, Banerjee, and
  Rao]{Madan_hydro1}
M.~Abhishek, A.~Dhanuka, D.~S. Banerjee and M.~Rao, \emph{Phys. Rev. E}, 2025,
  \textbf{112}, 045433\relax
\mciteBstWouldAddEndPuncttrue
\mciteSetBstMidEndSepPunct{\mcitedefaultmidpunct}
{\mcitedefaultendpunct}{\mcitedefaultseppunct}\relax
\EndOfBibitem
\bibitem[Roychowdhury \emph{et~al.}(2025)Roychowdhury, Dasgupta, and
  Rao]{Madan_hydro2}
A.~Roychowdhury, S.~Dasgupta and M.~Rao, \emph{Phys. Rev. E}, 2025,
  \textbf{112}, 044414\relax
\mciteBstWouldAddEndPuncttrue
\mciteSetBstMidEndSepPunct{\mcitedefaultmidpunct}
{\mcitedefaultendpunct}{\mcitedefaultseppunct}\relax
\EndOfBibitem
\bibitem[Gagnon \emph{et~al.}(2020)Gagnon, Dessi, Berezney, Boros, Chen, Dogic,
  and Blair]{gagnon_shear-induced_2020}
D.~A. Gagnon, C.~Dessi, J.~P. Berezney, R.~Boros, D.~T.-N. Chen, Z.~Dogic and
  D.~L. Blair, \emph{Physical Review Letters}, 2020, \textbf{125}, 178003\relax
\mciteBstWouldAddEndPuncttrue
\mciteSetBstMidEndSepPunct{\mcitedefaultmidpunct}
{\mcitedefaultendpunct}{\mcitedefaultseppunct}\relax
\EndOfBibitem
\bibitem[Broedersz and MacKintosh(2011)]{broedersz_molecular_2011}
C.~P. Broedersz and F.~C. MacKintosh, \emph{Soft Matter}, 2011, \textbf{7},
  3186--3191\relax
\mciteBstWouldAddEndPuncttrue
\mciteSetBstMidEndSepPunct{\mcitedefaultmidpunct}
{\mcitedefaultendpunct}{\mcitedefaultseppunct}\relax
\EndOfBibitem
\bibitem[Broedersz and MacKintosh(2014)]{broedersz_modeling_2014}
C.~Broedersz and F.~MacKintosh, \emph{Reviews of Modern Physics}, 2014,
  \textbf{86}, 995--1036\relax
\mciteBstWouldAddEndPuncttrue
\mciteSetBstMidEndSepPunct{\mcitedefaultmidpunct}
{\mcitedefaultendpunct}{\mcitedefaultseppunct}\relax
\EndOfBibitem
\bibitem[Fürthauer \emph{et~al.}(2021)Fürthauer, Needleman, and
  Shelley]{furthauer_design_2021}
S.~Fürthauer, D.~J. Needleman and M.~J. Shelley, \emph{New Journal of
  Physics}, 2021, \textbf{23}, 013012\relax
\mciteBstWouldAddEndPuncttrue
\mciteSetBstMidEndSepPunct{\mcitedefaultmidpunct}
{\mcitedefaultendpunct}{\mcitedefaultseppunct}\relax
\EndOfBibitem
\bibitem[Wyse~Jackson \emph{et~al.}(2022)Wyse~Jackson, Michel, Lwin, Fortier,
  Das, Bonassar, and Cohen]{wyse_jackson_structural_2022}
T.~Wyse~Jackson, J.~Michel, P.~Lwin, L.~A. Fortier, M.~Das, L.~J. Bonassar and
  I.~Cohen, \emph{Science Advances}, 2022, \textbf{8}, eabk2805\relax
\mciteBstWouldAddEndPuncttrue
\mciteSetBstMidEndSepPunct{\mcitedefaultmidpunct}
{\mcitedefaultendpunct}{\mcitedefaultseppunct}\relax
\EndOfBibitem
\bibitem[Xu and Safran(2015)]{Xu_PRE_2015}
X.~Xu and S.~A. Safran, \emph{Phys. Rev. E}, 2015, \textbf{92}, 032728\relax
\mciteBstWouldAddEndPuncttrue
\mciteSetBstMidEndSepPunct{\mcitedefaultmidpunct}
{\mcitedefaultendpunct}{\mcitedefaultseppunct}\relax
\EndOfBibitem
\bibitem[Murrell and Gardel(2012)]{Murrell_PNAS_2012}
M.~Murrell and M.~Gardel, \emph{Proc. Natl. Acad. Sci}, 2012, \textbf{109},
  20820--20825\relax
\mciteBstWouldAddEndPuncttrue
\mciteSetBstMidEndSepPunct{\mcitedefaultmidpunct}
{\mcitedefaultendpunct}{\mcitedefaultseppunct}\relax
\EndOfBibitem
\bibitem[Wang \emph{et~al.}(2014)Wang, Abhilash, Chen, Wells, and
  Shenoy]{Wang_BioPhysJ_2014}
H.~Wang, A.~Abhilash, C.~S. Chen, R.~G. Wells and V.~B. Shenoy,
  \emph{Biophysical Journal}, 2014, \textbf{107}, 2592--2603\relax
\mciteBstWouldAddEndPuncttrue
\mciteSetBstMidEndSepPunct{\mcitedefaultmidpunct}
{\mcitedefaultendpunct}{\mcitedefaultseppunct}\relax
\EndOfBibitem
\bibitem[Najma \emph{et~al.}(2024)Najma, Wei, Baskaran, Foster, and
  Duclos]{Najma_PNAS_2024}
B.~Najma, W.~Wei, A.~Baskaran, P.~Foster and G.~Duclos, \emph{Proc. Natl. Acad.
  Sci.}, 2024, \textbf{121}, year\relax
\mciteBstWouldAddEndPuncttrue
\mciteSetBstMidEndSepPunct{\mcitedefaultmidpunct}
{\mcitedefaultendpunct}{\mcitedefaultseppunct}\relax
\EndOfBibitem
\bibitem[Zakharov \emph{et~al.}(2024)Zakharov, Awan, Gopinath, Lee,
  Ramasubramanian, and Dasbiswas]{Zakharov_Sc_2024}
A.~Zakharov, M.~Awan, A.~Gopinath, S.~Lee, A.~Ramasubramanian and K.~Dasbiswas,
  \emph{Science Advances}, 2024, \textbf{10}, year\relax
\mciteBstWouldAddEndPuncttrue
\mciteSetBstMidEndSepPunct{\mcitedefaultmidpunct}
{\mcitedefaultendpunct}{\mcitedefaultseppunct}\relax
\EndOfBibitem
\bibitem[Farhadi \emph{et~al.}(2020)Farhadi, Ricketts, Rust, Das,
  Robertson-Anderson, and Ross]{farhadi_actin_2020}
L.~Farhadi, S.~N. Ricketts, M.~J. Rust, M.~Das, R.~M. Robertson-Anderson and
  J.~L. Ross, \emph{Soft Matter}, 2020, \textbf{16}, 7191--7201\relax
\mciteBstWouldAddEndPuncttrue
\mciteSetBstMidEndSepPunct{\mcitedefaultmidpunct}
{\mcitedefaultendpunct}{\mcitedefaultseppunct}\relax
\EndOfBibitem
\bibitem[Burla \emph{et~al.}(2019)Burla, Tauber, Dussi, van~der Gucht, and
  Koenderink]{burla_stress_2019}
F.~Burla, J.~Tauber, S.~Dussi, J.~van~der Gucht and G.~H. Koenderink,
  \emph{Nature Physics}, 2019, \textbf{15}, 549--553\relax
\mciteBstWouldAddEndPuncttrue
\mciteSetBstMidEndSepPunct{\mcitedefaultmidpunct}
{\mcitedefaultendpunct}{\mcitedefaultseppunct}\relax
\EndOfBibitem
\bibitem[Fürthauer \emph{et~al.}(2019)Fürthauer, Lemma, Foster, Ems-McClung,
  Yu, Walczak, Dogic, Needleman, and Shelley]{furthauer_self-straining_2019}
S.~Fürthauer, B.~Lemma, P.~J. Foster, S.~C. Ems-McClung, C.-H. Yu, C.~E.
  Walczak, Z.~Dogic, D.~J. Needleman and M.~J. Shelley, \emph{Nature Physics},
  2019, \textbf{15}, 1295--1300\relax
\mciteBstWouldAddEndPuncttrue
\mciteSetBstMidEndSepPunct{\mcitedefaultmidpunct}
{\mcitedefaultendpunct}{\mcitedefaultseppunct}\relax
\EndOfBibitem
\bibitem[Tanaka and Edwards(1992)]{tanaka_viscoelastic_1992}
F.~Tanaka and S.~F. Edwards, \emph{Macromolecules}, 1992, \textbf{25},
  1516--1523\relax
\mciteBstWouldAddEndPuncttrue
\mciteSetBstMidEndSepPunct{\mcitedefaultmidpunct}
{\mcitedefaultendpunct}{\mcitedefaultseppunct}\relax
\EndOfBibitem
\bibitem[Ronceray \emph{et~al.}(2016)Ronceray, Broedersz, and
  Lenz]{ronceray_fiber_2016}
P.~Ronceray, C.~P. Broedersz and M.~Lenz, \emph{Proceedings of the National
  Academy of Sciences}, 2016, \textbf{113}, 2827--2832\relax
\mciteBstWouldAddEndPuncttrue
\mciteSetBstMidEndSepPunct{\mcitedefaultmidpunct}
{\mcitedefaultendpunct}{\mcitedefaultseppunct}\relax
\EndOfBibitem
\bibitem[Nampoothiri \emph{et~al.}(2022)Nampoothiri, D'Eon, Ramola,
  Chakraborty, and Bhattacharjee]{nampoothiri_tensor_2022}
J.~N. Nampoothiri, M.~D'Eon, K.~Ramola, B.~Chakraborty and S.~Bhattacharjee,
  \emph{Physical Review E}, 2022, \textbf{106}, 065004\relax
\mciteBstWouldAddEndPuncttrue
\mciteSetBstMidEndSepPunct{\mcitedefaultmidpunct}
{\mcitedefaultendpunct}{\mcitedefaultseppunct}\relax
\EndOfBibitem
\bibitem[Lubensky \emph{et~al.}(2015)Lubensky, Kane, Mao, Souslov, and
  Sun]{lubensky_phonons_2015}
T.~C. Lubensky, C.~L. Kane, X.~Mao, A.~Souslov and K.~Sun, \emph{Reports on
  Progress in Physics}, 2015, \textbf{78}, 073901\relax
\mciteBstWouldAddEndPuncttrue
\mciteSetBstMidEndSepPunct{\mcitedefaultmidpunct}
{\mcitedefaultendpunct}{\mcitedefaultseppunct}\relax
\EndOfBibitem
\bibitem[Mao and Lubensky(2018)]{mao_maxwell_2018}
X.~Mao and T.~C. Lubensky, \emph{Annual Review of Condensed Matter Physics},
  2018, \textbf{9}, 413--433\relax
\mciteBstWouldAddEndPuncttrue
\mciteSetBstMidEndSepPunct{\mcitedefaultmidpunct}
{\mcitedefaultendpunct}{\mcitedefaultseppunct}\relax
\EndOfBibitem
\bibitem[Malakar \emph{et~al.}(2025)Malakar, Countryman, and
  Chakraborty]{Supplementary}
K.~Malakar, A.~Countryman and B.~Chakraborty, \emph{Supplemental Information},
  2025\relax
\mciteBstWouldAddEndPuncttrue
\mciteSetBstMidEndSepPunct{\mcitedefaultmidpunct}
{\mcitedefaultendpunct}{\mcitedefaultseppunct}\relax
\EndOfBibitem
\bibitem[Benoist \emph{et~al.}(2023)Benoist, Saggiorato, and
  Lenz]{Benoit_Lenz_2023}
F.~Benoist, G.~Saggiorato and M.~Lenz, \emph{Soft Matter}, 2023, \textbf{19},
  2970--2976\relax
\mciteBstWouldAddEndPuncttrue
\mciteSetBstMidEndSepPunct{\mcitedefaultmidpunct}
{\mcitedefaultendpunct}{\mcitedefaultseppunct}\relax
\EndOfBibitem
\bibitem[Ronceray and Lenz(2015)]{Ronceray_Lenz_2015}
P.~Ronceray and M.~Lenz, \emph{Soft Matter}, 2015, \textbf{11},
  1597--1605\relax
\mciteBstWouldAddEndPuncttrue
\mciteSetBstMidEndSepPunct{\mcitedefaultmidpunct}
{\mcitedefaultendpunct}{\mcitedefaultseppunct}\relax
\EndOfBibitem
\bibitem[Mao \emph{et~al.}(2010)Mao, Xu, and Lubensky]{mao_soft_2010}
X.~Mao, N.~Xu and T.~C. Lubensky, \emph{Physical Review Letters}, 2010,
  \textbf{104}, 085504\relax
\mciteBstWouldAddEndPuncttrue
\mciteSetBstMidEndSepPunct{\mcitedefaultmidpunct}
{\mcitedefaultendpunct}{\mcitedefaultseppunct}\relax
\EndOfBibitem
\bibitem[Zangwill(2013)]{Zangwill_EandM}
A.~Zangwill, \emph{Modern Electrodynamics}, Cambridge University Press,
  2013\relax
\mciteBstWouldAddEndPuncttrue
\mciteSetBstMidEndSepPunct{\mcitedefaultmidpunct}
{\mcitedefaultendpunct}{\mcitedefaultseppunct}\relax
\EndOfBibitem
\bibitem[Chakraborty \emph{et~al.}(2025)Chakraborty, Nampoothiri,
  Bhattacharjee, Chakraborty, and Ramola]{surajit_arxiv}
S.~Chakraborty, J.~N. Nampoothiri, S.~Bhattacharjee, B.~Chakraborty and
  K.~Ramola, \emph{Stress Response of Jammed Solids: Prestress and Screening},
  2025, \url{https://arxiv.org/abs/2509.14336}\relax
\mciteBstWouldAddEndPuncttrue
\mciteSetBstMidEndSepPunct{\mcitedefaultmidpunct}
{\mcitedefaultendpunct}{\mcitedefaultseppunct}\relax
\EndOfBibitem
\bibitem[Nampoothiri \emph{et~al.}(2020)Nampoothiri, Wang, Ramola, Zhang,
  Bhattacharjee, and Chakraborty]{nampoothiri_emergent_2020}
J.~N. Nampoothiri, Y.~Wang, K.~Ramola, J.~Zhang, S.~Bhattacharjee and
  B.~Chakraborty, \emph{Physical Review Letters}, 2020, \textbf{125},
  118002\relax
\mciteBstWouldAddEndPuncttrue
\mciteSetBstMidEndSepPunct{\mcitedefaultmidpunct}
{\mcitedefaultendpunct}{\mcitedefaultseppunct}\relax
\EndOfBibitem
\bibitem[Ting and Chen(2005)]{ting_poissons_2005}
T.~C.~T. Ting and T.~Chen, \emph{The Quarterly Journal of Mechanics and Applied
  Mathematics}, 2005, \textbf{58}, 73--82\relax
\mciteBstWouldAddEndPuncttrue
\mciteSetBstMidEndSepPunct{\mcitedefaultmidpunct}
{\mcitedefaultendpunct}{\mcitedefaultseppunct}\relax
\EndOfBibitem
\bibitem[Shivers \emph{et~al.}(2020)Shivers, Arzash, and
  MacKintosh]{shivers_nonlinear_2020}
J.~L. Shivers, S.~Arzash and F.~MacKintosh, \emph{Physical Review Letters},
  2020, \textbf{124}, 038002\relax
\mciteBstWouldAddEndPuncttrue
\mciteSetBstMidEndSepPunct{\mcitedefaultmidpunct}
{\mcitedefaultendpunct}{\mcitedefaultseppunct}\relax
\EndOfBibitem
\bibitem[Berezney \emph{et~al.}(2022)Berezney, Goode, Fraden, and
  Dogic]{berezney_extensile_2022}
J.~Berezney, B.~L. Goode, S.~Fraden and Z.~Dogic, \emph{Proceedings of the
  National Academy of Sciences}, 2022, \textbf{119}, e2115895119\relax
\mciteBstWouldAddEndPuncttrue
\mciteSetBstMidEndSepPunct{\mcitedefaultmidpunct}
{\mcitedefaultendpunct}{\mcitedefaultseppunct}\relax
\EndOfBibitem
\end{mcitethebibliography}
\bibliographystyle{rsc} 

\section*{Supplemental Material}

\subsection*{Simulation details}
We initiate our simulation with an unperturbed network of 2500 springs: $l_{ij}=l_0$ and $\theta_{ijk}=0$ for all springs. We then apply a force dipole of strength $|F|_{0}$, along the horizontal axis on a pair of nearest neighbor junctions.
We minimize the energy of the network in the presence of this  force dipole, and record this configuration and its energy. The final configuration is obtained by iteratively solving the equations of motion of every node in the overdamped limit, until force balance is achieved for all individual nodes 
The final configuration is obtained by iteratively solving the equations
of motion of every node in the overdamped limit, until the total energy of the system changes by an amount less than 0.0001 from one time step to the next. We then increase $|F|$ by an amount $\Delta |F|$, and repeat the procedure.  This ``annealing'' procedure provides us with the set of configurations as a function of $|F$, which are analyzed to obtain the results discussed in the following sections.
    In our simulations, $l_0=2$, $|F|_0=\Delta |F|$, and  $\Delta |F| = 2\kappa/l_0$~. The network has 58 columns and 50 rows, i.e. 2900 junctions, 8700 springs, and 8700 midpoints.

\subsection{Kagome construction}
Even though it feels intuitive to look for patterns in  buckling angles to characterize such changes, our analysis did not identify any clear signatures. 
We use a different construction, which reveals distinct patterns of buckling in the three phases.
It is important to point out that the Kagome structure  is a theoretical construct that is useful for analyzing buckling patterns: in our simulation model,  there are no springs connecting the Kagome points.
The Kagome network consists of corner-sharing triangles as shown in Fig. \ref{kagome}B.  The soft modes  
involve coordinated counterrotations of corner connected triangles~\cite{lubensky_phonons_2015,mao_maxwell_2018} as depicted by the arrows in Fig. \ref{kagome}B.  To characterize the buckling patterns,
we take a pair of corner-sharing up and down triangles, shown in red (or blue) in figure \ref{kagome}B, and measure the rotation of each of the two component triangles.
There can be three different outcomes: 
(i) the triangles have not rotated with respect to the initial unperturbed state, 
(ii) both triangles have rotated in the same direction (either clockwise or anticlockwise) with respect to the initial configuration, or
(iii) the top and bottom triangles are rotated in opposite directions forming a twisted unit, which is the  basic deformation characterizing the Kagome soft modes. 
For triangles falling into category (iii), we assign colors red and blue to these units depending on the direction of the twist, as shown in Fig. \ref{kagome}B.
We calculate the twist of every pair of triangles in the network, and plot the location of each twisted unit with a color (blue or red) describing its twist (Fig. \ref{kagome}C).
The locations where the deformation is of type (i) or (ii) are left empty. This color mapping thus provides a visual representation of the projection of the buckling patterns on to the soft modes of a Kagome lattice.

\subsection*{Stress response to a single force}
Explicitly, the Fourier transform of the stress response, $\hat{\tilde{\sigma}}({\bf q})$ to a single force can be written as $\tilde{\sigma}_{\alpha \beta}({\bf q}) = \tilde{G}_{\alpha \beta \gamma}({\bf q}) \tilde{F}_{\gamma}({\bf q})$, where $\tilde{G}$ is the Green function in the Fourier space.
\begin{align}
    \tilde{G}_{\alpha \beta \gamma}({\bf q}) = \tilde{\mathscr{G}}_{\alpha \beta \gamma}({\bf q})  + \nu (|F|) \left[ \tilde{\mathscr{G}}_{\alpha \beta \gamma}({\bf q})  + \frac{i \delta_{\alpha \beta }q_{\gamma} - i \delta_{\alpha \gamma}q_{\beta} - i \delta_{ \beta \gamma}q_{\alpha}}{q^2} \right] ~,
\end{align}
where  $\tilde{\mathscr{G}}_{\alpha \beta \gamma}({\bf q})  = \frac{i }{q^4}[ q_{\alpha}q^2 \delta_{\beta \gamma} + q_{\beta}q^2 \delta_{\alpha \gamma} -q_{\alpha}q_{\beta}q_{\gamma} ]$.
Adding the responses to a forces of magnitude $|F| \hat{x}$ at $(d, 0)$, and  $-|F| \hat{x}$$(-d, 0)$, produces the response to an extensile dipole of length $2d$ located at the origin.

\subsection*{Independent components of stress tensor in polar coordinates}
The three independent polar components of the real-space stress tensor are related to the cartesian components by:
\begin{align}
\sigma_{rr} =&  \frac{1}{2} (\sigma_{xx} - \sigma_{yy}) \cos{2 \theta} + \sigma_{xy} \sin{2\theta} + \frac{1}{2}(\sigma_{xx}+\sigma_{yy}) \label{rrResponse}\\
  \sigma_{\theta \theta} =&  \frac{1}{2} (\sigma_{yy} - \sigma_{xx}) \cos{2 \theta} - \sigma_{xy} \sin{2\theta} + \frac{1}{2}(\sigma_{xx}+\sigma_{yy}) \label{ttResponse}\\
  \sigma_{r \theta} =&  \frac{1}{2} (\sigma_{yy} - \sigma_{xx}) \sin{2 \theta} + \sigma_{xy} \cos{2\theta} \label{rtResponse}
\end{align}
where $\theta$ is the polar angle, and $\sigma_{xx}, \sigma_{yy}, \sigma_{xy}$ are the cartesian components.

\subsection*{Fitting Procedure}

Fitting of Fourier-space stress response functions was performed following a detailed procedure of re-gridding and re-binning of stress data. Data is first centered such that the force dipole lies at $\vec{r}=0$, then a standard Discrete Fourier Transform (DFT) is applied. Given that the nearly lies on a triangular lattice, the wave vectors at which the DFT is evaluated are spaced by $2\pi/L_x$ in the $q_x$ direction (up to extrema of $\pm \pi/d$) and $2\pi/L_y$ in the $q_y$ direction (up to extrema of $\pm2\pi/(\sqrt{3}d)$). 

While the above process yields a spectrum $\sigma_{ij}(\vec{q})$ measured on a Cartesian rectangular lattice, theoretical predictions for the stress response are most naturally expressed in polar coordinates. The theoretical forms of $\sigma_{ij}(\vec{q})$ are all of the form $f(\theta)\sin(dq_x/2)/q$ (note that $f(\theta)=qG_{ijx}(q,\theta)$, the form of which is determined by the elastic modulus tensor), which indeed matches the measured forms of $\sigma_{ij}(q)$ near $q=0$. In order to reduce the dimensionality of the fit to only take into account angular variation, the data is multiplied by $q/\sin(dq_x/2)$ such that $f(\theta)$ can be fit to yield the elastic modulus tensor.

Over the range of $q$ where $\sigma_{ij}(\vec{q})\sim  \sin(dq_x/2)/q$, the measured $f(\theta)$ is averaged into 30 angular bins. The one-dimensional dataset can now be fit to the VCTG theoretical form of $f(\theta)$. This data was captured successfully by the form of $f(\theta)$ corresponding to an isotropic elastic modulus tensor, the simplest possible elastic modulus for an isotropic material. The corresponding form of $f(\theta)$ depends only upon a dimensionless combination of the Lamé parameters $\lambda,\mu$, which can be taken to be Poisson's ratio $\nu$. 

There are three components of stress to fit in 2D, and all ought to be consistent with the same elastic modulus tensor. The Python package SymFit is used to simultaneously fit all components of $\sigma_{ij}(\vec{q})$ to the same value of $\nu$, constrained to a range from $-1$ to $1$ in 2D. The outcome of this fitting procedure yields the values of $\nu$ reported in Table 1. A table of corresponding $r^2$ values for these fits can be seen below.

\begin{table}[h]
 \begin{center}
  \begin{tabular}{ |c|c|c|c|c|c|c| }
     \hline
     F& 5 & 10 & 15 & 20 & 25 & 30 \\\hline 
     $r^2$  & 0.9905 & 0.9948 & 0.9921 & 0.9906 & 0.9901 & 0.9887   \\
     \hline
  \end{tabular}\caption{\footnotesize{Table of $r^2$ values obtained by fitting stress response to the VCTG theory.}}
  \label{r2 values}
 \end{center}
\end{table}

\subsection*{Comparing simulations to VCT predictions}

In order to test the validity of this approach, we use the computed values of $\nu$ to compute $\sigma_{\theta \theta}$ at $F=10$ and $F=30$ in fig \ref{fig:stressTT}C and  fig \ref{fig:stressTT}D, respectively. 
These figures show that the non-linear dielectric generalization of VCT can indeed reproduce the stress patterns observed in simulations, semi quantitatively.
The full nonlinear response of the elastic network in all three mechanical regimes, L (linear) , N (non-linear),  and  R (rectified), is captured by just a renormalization of the Poisson ratio.  This is a remarkable result since VCT is a continuum theory in which the microstructural changes such as buckling are represented via a single tensor, $K$.   The correlation of the discontinuous changes in $\nu(F)$ with the distinct shift in the buckling patterns demonstrates that this fundamental premise of the VCTG framework is borne out in this system. 

\subsection*{Response of a disordered network}

\begin{figure}[h!]
\centering
\includegraphics[width = 0.45\textwidth]{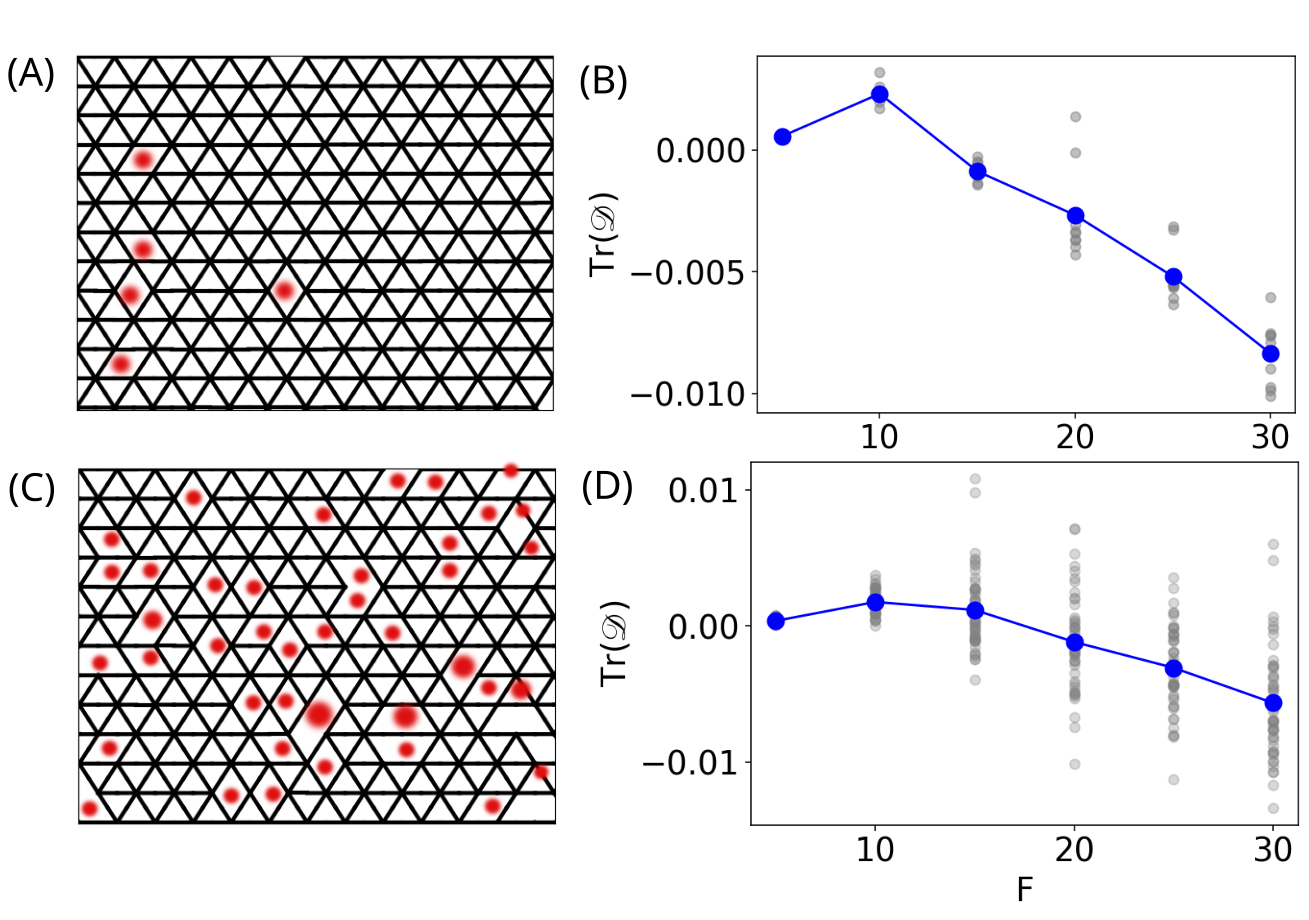}
\caption{ Disordered networks with (A) $1\%$ dilution and (C) $10\%$ dilution. The sites missing bonds are marked with red circles.  (B) and (D) show far-field stress response at different value of applied extensile force dipole corresponding to (A) and (C) respectively. Gray dots are stress values from individual simulations and blue line is disorder averaged over 10 (1\% dilution) and 50 (10\% dilution) realizations.}
\label{fig:disordered_network}
\end{figure}

\begin{figure}[h!]
\centering
\includegraphics[width = 0.45\textwidth]{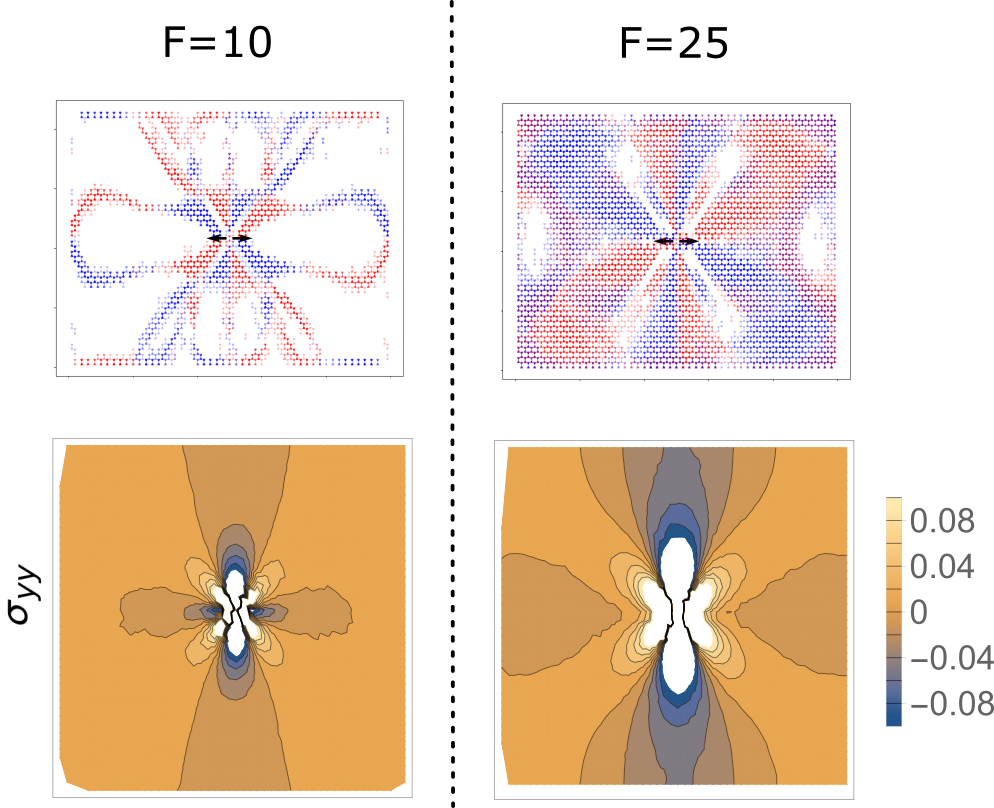}
\caption{Buckling pattern (top row) and stress patterns (bottom row) of 10\% diluted lattice in phase L (F=10) and phase R (F=25), ensemble averaged over 10 realizations.}
\label{fig:disorder-kagome-stress}
\end{figure}
In order to check the robustness of the mechanical screening picture, we analyze the  effects of a force dipole in a non-linear, diluted triangular lattice.
Random bonds in the network are removed to create a disordered system.
A force dipole acts on two centrally located nodes as before.
Fig \ref{fig:disordered_network} shows two networks with different dilutions.
(A) has 1\% bonds removed and (C) has 10\% bonds removed.
The figures to the right of the networks show the rectification curve for the corresponding dilution ensemble averaged over multiple realizations.

\subsection*{Discussion on Poisson ratio}
The field $\hat{E}$, which has the same dimensions as $\hat{\sigma}$, is the field that obeys the compatibility relation normally associated with the strain tensor in classical elasticity theory~\cite{nampoothiri_tensor_2022}. $\hat{E}$ can be related to relative strains between two arbitrary configurations via a dimension-full constant~~\cite{surajit_arxiv}. Therefore, $K$ is also related to elastic moduli obtained from strain-based measurements via dimensionfull constants. Notably, the Poisson ratio is dimensionless and, therefore, can be computed within the VCT framework.

As seen from Table ~\ref{nonlinear}, the Poisson ratio approaches unity with increasing $|F|$.  In 2D,  $\nu = (\lambda-\mu)/(\lambda +\mu)$, where $\lambda$ is the bulk modulus and $\mu$, the shear modulus.
Therefore,  $\nu \rightarrow 1$ would imply that the shear modulus $\mu \rightarrow 0$, which indicates that the system is on the verge of becoming unstable to shear. Since we cannot deduce the shear modulus from the stress response, we cannot provide an explicit check of this instability.  We speculate that the rectfication phase, (R), characterized by a a contractile response to an extensile force dipole is a precursor to the initiation of an activity driven plasticity or flow~\cite{berezney_extensile_2022}. 

\end{document}